\documentclass[letterpaper,11pt]{article}
\usepackage[utf8]{inputenc}

\usepackage{tabularx}

\usepackage{amsmath, amssymb}
\usepackage{amsxtra}

\usepackage{bigints}
\usepackage{float}
\usepackage{graphicx}

\usepackage[margin=1in,letterpaper]{geometry}
\usepackage{subcaption}

\usepackage[final]{hyperref}

\hypersetup{
	colorlinks=true,       % false: boxed links; true: colored links
	linkcolor=blue,        % color of internal links
	citecolor=blue,        % color of links to bibliography
	filecolor=magenta,     % color of file links
	urlcolor=blue         
}

\usepackage{authblk}

\title{Dynamics and merger rate of primordial black holes in a cluster}

\author{
    Viktor D. Stasenko$^{1,2, }$\thanks{\href{StasenkoVD@gmail.com}{StasenkoVD@gmail.com}}
    }

\author{
    Alexander A. Kirillov$^{1, }$\thanks{\href{AAKirillov@mephi.ru}{AAKirillov@mephi.ru}} 
    }

\author{
    Konstantin M. Belotsky$^{1, }$\thanks{\href{KMBelotskij@mephi.ru}{KMBelotskij@mephi.ru}} 
    }

\affil{
    $^1$National Research Nuclear University MEPhI
    (Moscow Engineering Physics Institute), 
    \par
    115409 Kashirskoe shosse 31, Moscow, Russia
    
    ~
    
    $^2$Institute of Physics, Southern Federal University,
    \par
    Stachki 194, Rostov on Don 344090, Russia
    
    }

\date{}

\usepackage[
    style=phys, 
    biblabel = brackets,
    eprint = true,
    language=english,
    autolang=other, 
    sorting=none, 
    bibencoding=auto, 
    backend=biber,
    maxbibnames=3,
    doi=false, 
    url=false]{biblatex}
\bibliography{bib}
\AtEveryBibitem{
	\clearfield{day}
	\clearfield{month}
	\clearfield{endday}
	\clearfield{endmonth}
	\clearfield{eid}
}

\newcommand{\D}{\mathrm{d}\,\!}

\begin{document}

\maketitle

\begin{abstract}
    The PBHs clusters can be source of gravitational waves, and the merger rate depends on the spatial distribution of PBHs in the cluster which changes over time. It is well known that gravitational collisional systems experience the core collapse, that leads to significantly increase of the central density and shrinking of the core. After core collapse, the cluster expands almost self-similarly (i.e. density profile extends in size without changing its shape). These dynamic processes affect at the merger rate of PBHs. In this paper the dynamics of the PBH cluster is considered using the Fokker-Planck equation. We calculate the merger rate of PBHs on cosmic time scales and show that its time dependence has a unique signature. Namely, it grows by about an order of magnitude at the moment of core collapse which depends on the characteristic of the cluster, and then decreases according to the dependence $\mathcal{R} \propto t^{-1.48}$. It was obtained for monochromatic and power-law PBH mass distributions with some fixed parameters. Obtained results can be used to test the model of the PBH clusters via observation of gravitational waves at high redshift.
\end{abstract}

\section{Introduction}

The hypothesis of the formation of primordial black holes (PBHs) in the early Universe was proposed in the works \cite{1967SvA....10..602Z, 1971MNRAS.152...75H, 1974MNRAS.168..399C}. Nowadays, there are a lot of mechanisms leading to the production of PBHs \cite{Dolgov:1992pu, Garcia-Bellido:1996mdl, 1999PhRvD..59l4014J, 2010RAA....10..495K, 2013PhRvD..87f3519K, Garcia-Bellido:2017mdw, 2017PhRvL.119c1103C, 2019EPJC...79..246B, 2019JCAP...10..077C, 2020PhRvL.125r1304K}. Historically, PBHs have attracted an interest due to the dark matter (DM) problem. However, the modern observational data restrict the possibility of PBHs to provide all the dark matter \cite{Carr:2020xqk}. The PBHs with masses $\sim 10^{-16} \div 10^{-10} \, M_{\odot}$ can explain all DM \cite{2021JPhG...48d3001G, 2020PhRvL.125r1304K}. 
Nevertheless, the PBHs hypothesis might have various astrophysical manifestations, such as gravitational waves (GWs) from black holes merging, nature of high redshift quasars, production of DM halos, etc.
The registration of GWs by LIGO/Virgo \cite{LIGOScientific:2016aoc} has inspired a renewed interest to PBHs \cite{2016PhRvL.116t1301B, 2016JCAP...11..036B, 2017PDU....15..142C, 2017JCAP...09..037R, 2018PDU....22..137C, 2019JCAP...02..018R, PhysRevD.96.123523, 2021JCAP...04..031K, 2021arXiv211009509S} because the GWs events are difficult to explain by stellar origin \cite{2020JCAP...12..017D}. The PBHs could be responsible for the formation of early DM halos \cite{2019PhRvD.100h3528I} at $z > 10$ which explain the near-IR cosmic infrared background fluctuations \cite{Kashlinsky:2016sdv}. In addition, early quasars founded at high redshifts \cite{2018Natur.553..473B, 2019ApJ...872L...2M, 2020ApJ...897L..14Y} might have a primordial origin \cite{PhysRevD.90.083514, 2017PASA...34...31V, 2020ARA&A..58...27I}. Moreover, mergers of black holes in a relativistic PBH cluster could solve ``$H_0$-tension'' problem \cite{2021PDU....3200833E}.

The question of PBHs clustering might play a decisive role in observational effects.
{It is known that various models predict formation of PBH clusters both in the postinflation epoch (an initial clustering)} \cite{Rubin:2000dq, 2001JETP...92..921R, 2005APh....23..265K, 2019PhRvD.100j3003D, 2019PhRvD.100l3544M, 2020JCAP...03..004Y, 2021arXiv210703580K} {and in the early epoch of structure formation due to Poisson fluctuations} \cite{2003ApJ...594L..71A, 2019PhRvD.100h3528I, 2020JCAP...09..022J, 2020JCAP...11..028D}. 
This study considers the model of cluster formation due to collapse of domain walls produced as a result of quantum fluctuations of scalar field(s) at the inflation stage \cite{Rubin:2000dq, 2001JETP...92..921R, 2005APh....23..265K}. {Following this model, a falling power-law mass distribution is assumed here in a wide PBH mass range.} The cluster decouples from the Hubble flow and forms a virialized system at the redshift $z_f \sim 10^4$ \cite{2001AstL...27..759D, 2019EPJC...79..246B}. The parameters of the resulting cluster depend on the specific model of scalar field potential. Typically, it has characteristics close to globular star clusters with a wide range of PBH masses. However, the observed signatures (e.g., the merger rate) of the PBHs cluster depends on its structure at a specific redshift. Hence, the cluster evolution is important. The dynamics of a PBHs cluster was considered in the works \cite{2019JCAP...02..018R, 2021Univ....7...18T, 2020MNRAS.496..994K} with help of the N-body simulation. However, such calculations are only suitable for clusters with a small number of black holes.
In this work, one uses the Fokker--Planck equation to study dynamics of the PBHs cluster. This approach allows practically studying an arbitrary range of the possible cluster parameters in a short computational time. 

Additionally, note that the picture can be changed by including other processes. It might be caused by taking into account baryons or/and the dark matter which change the dynamics of the cluster directly or through accretion processes.
However, these effects are the question of a separate investigation. Here, the effect of the pure gravitational evolution of the PBH cluster is considered.

In this paper, we show that during the evolution, the spatial distribution of PBHs is changed due to two-body relaxation processes and the mass segregation. We estimate the time dependence of the PBH merger rate inside the cluster and show that it does not depend on the initial parameters of the cluster and has the unique signature. The obtained results might be used to test the model of the PBH cluster with the help of a future generation of gravitational waves detectors which will be able to detect black holes mergers at high redshifts.
The effect under consideration is one of the rare possibly observable effects inherent in models of PBH clusters that may include other effects such as gravitational microlensing on the cluster \cite{Toshchenko:2019mth}. {In addition, effects giving constraints on PBHs as DM and constraints themselves in case of PBH clustering should be reconsidered} \cite{2019EPJC...79..246B}.

\section{The Fokker-Planck equation approach}

The orbit-averaged Fokker-Planck (FP) equation is often used to study evolution of globular star clusters \cite{1979ApJ...234.1036C, 1980ApJ...242..765C, 1989ApJ...343..725Q, 1990ApJ...351..121C} and galactic nuclei \cite{1991ApJ...370...60M, 2017ApJ...848...10V, 2009ApJ...694..959M, 2015ApJ...804...52M}. In our case, this equation could be used to describe time evolution of the distribution function of PBHs $f(E)$ as a result of diffusion in energy space. Note, the distribution function $f(E)$ depends only on energy (per unit mass) $E = v^2/2 + \phi(r)$. The FP equation for a multi-mass cluster is given by \cite{2017ApJ...848...10V}
\begin{equation} \label{fp_E}
    4 \pi^2 p(E) \frac{\partial f_i}{\partial E} = - \frac{\partial}{\partial E} \left ( m_i D_E f_i + D_{EE} \frac{\partial f_i}{\partial E} \right) - \nu f_i,
\end{equation}
where the index $i$ refers to the $i$-th mass type of PBHs, the last loss-cone term $\nu f_i$ describes the absorption of PBHs by the massive central black hole (CBH) \cite{2017ApJ...848...10V}, and the coefficients $D_{E}$ and $D_{EE}$ are
\begin{align}
    D_{E} &= -16 \pi^3 \Gamma \sum_j m_j \int_{\phi (0)}^E \D{E'} \, f_j(E') p (E'), 
    \label{D_E} 
    \\
    D_{EE} &= - 16 \pi^3 \Gamma \sum_j m_j^2 \left ( q(E) \int_E^0 \D{E'} \, f_j(E') + \int_{\phi(0)}^{E} \D{E'} \, q(E') f_j(E') \right) 
    \label{D_EE},
\end{align}
where the sum is over all the types of PBH. $\Gamma = 4 \pi G^2 \ln{\Lambda}$, $\ln{\Lambda}$ is the Coulomb logarithm, and $G$ is the Newtonian gravitational constant. The expressions for $q(E)$ and $p(E)$ are 
\begin{align}
    q(E) &= \frac{4}{3} \int_{0}^{\phi^{-1}(E)} \D{r} \, r^2 \Big[ 2 \big( E - \phi(r) \big) \Big]^{3/2} 
    \label{qE}, 
    \\
    p(E) &= 4 \int_{0}^{\phi^{-1}(E)} \D{r} \, r^2 \sqrt{2 \big( E - \phi(r) \big)} 
    \label{pE}, 
\end{align}
where $\phi^{-1}(E)$ is the root of the equation $E = \phi(r)$. $\phi(r)$ is the gravitational potential which is given by Poisson equation solution
\begin{equation} 
    \label{grav_potential}
    \phi(r) = -4 \pi G \left ( \frac{1}{r} \int_0^r \, \D{r'} \, r'^2  \rho(r') + \int_r^{\infty} \D{r'} \, r' \rho(r')  \right) - \frac{G M_{\bullet}}{r}.
\end{equation}
The last term is the potential of the CBH with the mass $M_{\bullet}$, and $\rho(r)$ is the density profile
\begin{equation} 
    \label{rho}
    \rho(r) = 4 \pi \sum_{j} \int_{\phi(r)}^{0} \D{E} \, f_j(E) \sqrt{2 \big( E - \phi(r) \big)}.
\end{equation}

The evolution of cluster is described by both Fokker-Planck equation \eqref{fp_E} and Poisson equation \eqref{grav_potential}. The diffusion coefficients \eqref{D_E} and \eqref{D_EE} are calculated via distribution function from a previous time step; therefore, \eqref{fp_E} is solved as linear equation at a time step $\Delta \tau$. Then, \eqref{grav_potential} and \eqref{rho} are solved by iterations, and so on. This calculation algorithm can be found in more detail in the papers \cite{1980ApJ...242..765C, 2017ApJ...848...10V}. 

The evolution of PBH cluster is similar to the dynamics of star clusters and is driven by two-body relaxation. This leads to the effect of a core collapse  \cite{1980ApJ...242..765C}, as a result of which the radius and the mass of the core (the central part of the cluster) shrink to zero $r_c \rightarrow 0$, $M_c \rightarrow 0$, while the density of core goes to infinity $\rho_c \rightarrow \infty$. 
This phenomena is also known as ``gravothermal catastrophe'' \cite{1968MNRAS.138..495L}. 
The time dependence of $r_{c}(t)$ and $M_{c}(t)$ are given by \cite{1987degc.book.....S}
\begin{align}
    r_{c}(t) & = r_{c}(0) \left ( 1 - \frac{t}{t_{cc}} \right)^{0.53} \label{r_c}, 
    \\
    M_{c}(t) & = M_{c}(0) \left (1 - \frac{t}{t_{cc}} \right)^{0.42} \label{M_c},
\end{align}
where $t_{cc}$ is the core collapse time
\begin{equation} 
    \label{t_cc}
    t_{cc} \sim 30 \left ( \frac{r_c (0)}{1 \, \text{pc}} \right)^{3/2} \left ( \frac{M_{c} (0)}{10^5 \, M_{\odot}} \right)^{1/2} \left (\frac{10 \, M_{\odot}}{m} \right) \, \text{Myr}.
\end{equation}

Note, Eqs.~$\eqref{r_c}$ and $\eqref{M_c}$ do not have a physically reasonable limit at $t \rightarrow t_{cc}$. In real stellar clusters, a core collapse stops due to the following reasons: they are formation of binary stars \cite{1975MNRAS.173..729H, 1992PASP..104..981H, 2012MNRAS.425.2493B}, or the presence of primordial binaries \cite{1989Natur.339...40G, 2006MNRAS.368..677H}, or the influence of a massive CBH \cite{2017ApJ...848...10V, 2018PhRvD..98b3021S} which acts as a heating source. After the termination of the core collapse, cluster enters to a nearly self-similar expansion stage.

\section{The merging of primordial black holes}

The PBH cluster is the great place for intense black holes mergers. The cross section of binary PBH formation due to emission of gravitational waves is given by \cite{1989ApJ...343..725Q, 2002ApJ...566L..17M}
\begin{equation}
    \sigma = 2 \pi \left ( \frac{85 \pi}{6 \sqrt{2}} \right)^{2/7} \frac{G^2 (m + m')^{10/7} m^{2/7} m'^{2/7}}{c^{10/7} v_\text{rel}^{18/7}},
\end{equation}
where $v_\text{rel}$ is the relative velocity of PBHs, $m$ and $m'$ are the masses of PBHs, and $c$ is the speed of light. It is typically assumed that the forming binary is immediately \mbox{merged~\cite{1989ApJ...343..725Q, 2016PhRvL.116t1301B, 2017PDU....15..142C, 2021Univ....7...18T, 2021arXiv210911376G}}. Possible formation of binary systems due to triple PBH interactions are not taken into account here and commented in the end of this section. The merger rate of black holes per cluster will be 
\begin{equation} 
    \label{mr_1}
    \mathcal{R} = \frac{4 \pi}{m^2} \int \D{r} \, r^2 \rho^2 \sigma v_\text{rel},
\end{equation}
 where $\rho$ is the density of PBHs. It is assumed for the sake of simplicity of calculations that all PBHs in the cluster have the same mass $m$. For estimation, one can write that inside the cluster core $\rho(r) = \rho_{c}$ and $v_\text{rel} = \sqrt{G M_c / r_c}$, where $M_c$ and $r_c$ are the mass and the radius of the core, respectively. Outside the core, the density falls as $\rho \propto r^{-\beta}$ ($\beta \approx$ 2.2 is the typical value for a globular star cluster). Then, the merger rate can be written as
\begin{equation} 
    \label{mr_2}
    \mathcal{R} \sim \frac{4 \pi r_c^3}{3 m^2} \, \rho_c^2 \sigma \sqrt{G M_c / r_c} \sim 10^{-9} \left ( \frac{M_c}{10^5 \, M_{\odot}} \right)^{17/14} \left ( \frac{r_c}{1 \, \text{pc}} \right)^{-31/14} \, \text{yr}^{-1}.
\end{equation}
Note, the estimation \eqref{mr_2} depends only on the parameters of the central part of the cluster, even if the total mass of the cluster is much greater than the core mass.

Using the expressions \eqref{r_c} and \eqref{M_c}, one can get the time evolution of the merger rate:
\begin{align} \label{mr_time}
    \mathcal{R} (t) = \mathcal{R} (0) \left ( 1 - \frac{t}{t_{cc}} \right)^{-0.66}.
\end{align}

Moreover, here we assume that the clusters contain central massive black hole. Then, an additional channel for black hole mergers is added, namely the capture of less massive PBHs by the CBH. The growth rate of the CBH mass can be obtained as
\begin{equation} 
    \label{cbh_mr_1}
    \dot{M_{\bullet}} \sim \rho \sigma_{c} v_\text{rel} \sim 10^{-11} \left (\frac{r_{c}}{1 \, \text{pc}} \right)^{-5/2} \left ( \frac{M_{c}}{10^5 \, M_{\odot}} \right)^{1/2} \left ( \frac{M_{\bullet}}{10^3 \, M_{\odot}} \right)^{2} \, M_{\odot} \, \text{yr}^{-1} ,
\end{equation}
where $\sigma_c = 4 \pi r_g^2 (c / v_\text{rel})^2$ is the capture cross section of particles by the CBH \cite{1975ctf..book.....L} and $r_{g}$ is the gravitational radius of CBH. Using \eqref{r_c} and \eqref{M_c}, one can get
\begin{equation} \label{cbh_rate_time}
    \frac{\dot{M_{\bullet}}}{M_{\bullet}^2} \propto \left ( 1 - \frac{t}{t_{cc}} \right)^{-1.12}. 
\end{equation}
Here, both \eqref{mr_time} and \eqref{cbh_rate_time} diverge at $t \rightarrow t_{cc}$, that is not a physical result. As noted above, when  the core collapse is terminated in real systems,  the merger rate decreases due to expansion of the cluster.
However, one should keep in mind that the presented analysis is simplified and does not take into account the presence of a PBH mass spectrum in a cluster. The calculations performed for the cluster with the PBH mass spectrum are presented in this study.
Nevertheless, this result 
can be useful for 
probing the model of PBHs clusters 
if an enhancement of the merger rate will be observed. Observation of an enhancement of a merger rate through gravitational waves with increasing redshift would support PBH cluster hypothesis.
The calculation details of the merger rate for the PBH cluster with a spectrum mass is described in \cite{2021Physi...3..372S}.

It is important to note that PBH binaries can be also formed through three-body interactions \cite{1975MNRAS.173..729H}. However, the evolution of such binary systems cannot be studied within the framework of the Fokker-Planck equation. These binaries can both be destroyed in a cluster and become tighter systems with subsequent merging through emission of gravitational waves. The study of these processes requires N-body simulation and is outside the scope of this work.

\section{Evolution of PBHs clusters}

\subsection{PBHs clusters with monochromatic mass spectra}

In order to show main features of the cluster dynamics, we first consider the simplified model. In this section, the masses of all PBHs are equal to $M_{\odot}$, the CBH mass is $100 \, M_{\odot}$, and the total mass of the cluster is $10^5 \, M_{\odot}$. We take the initial density profile of PBHs in the cluster in the following form
\begin{equation} 
    \label{PBH_profile}
    \rho(r) = \rho_0 \left ( \frac{r}{r_0} \right)^{-\gamma} \left [ 1 +  \left (\frac{r}{r_0} \right)^2 \right]^{\frac{\gamma - \beta}{2}},
\end{equation}
where $\rho_0$ is the normalization constant. Note, that initial cusp $\rho \propto r^{-\gamma}$ is inevitable due to presence of CBH. The cusp must be steeper than $r^{-1/2}$, otherwise the initial distribution will not be physical \cite{2013degn.book.....M}.

\begin{figure}[!th]
    \centering
    \begin{subfigure}[t]{0.48\linewidth}
    \includegraphics[width=\linewidth]{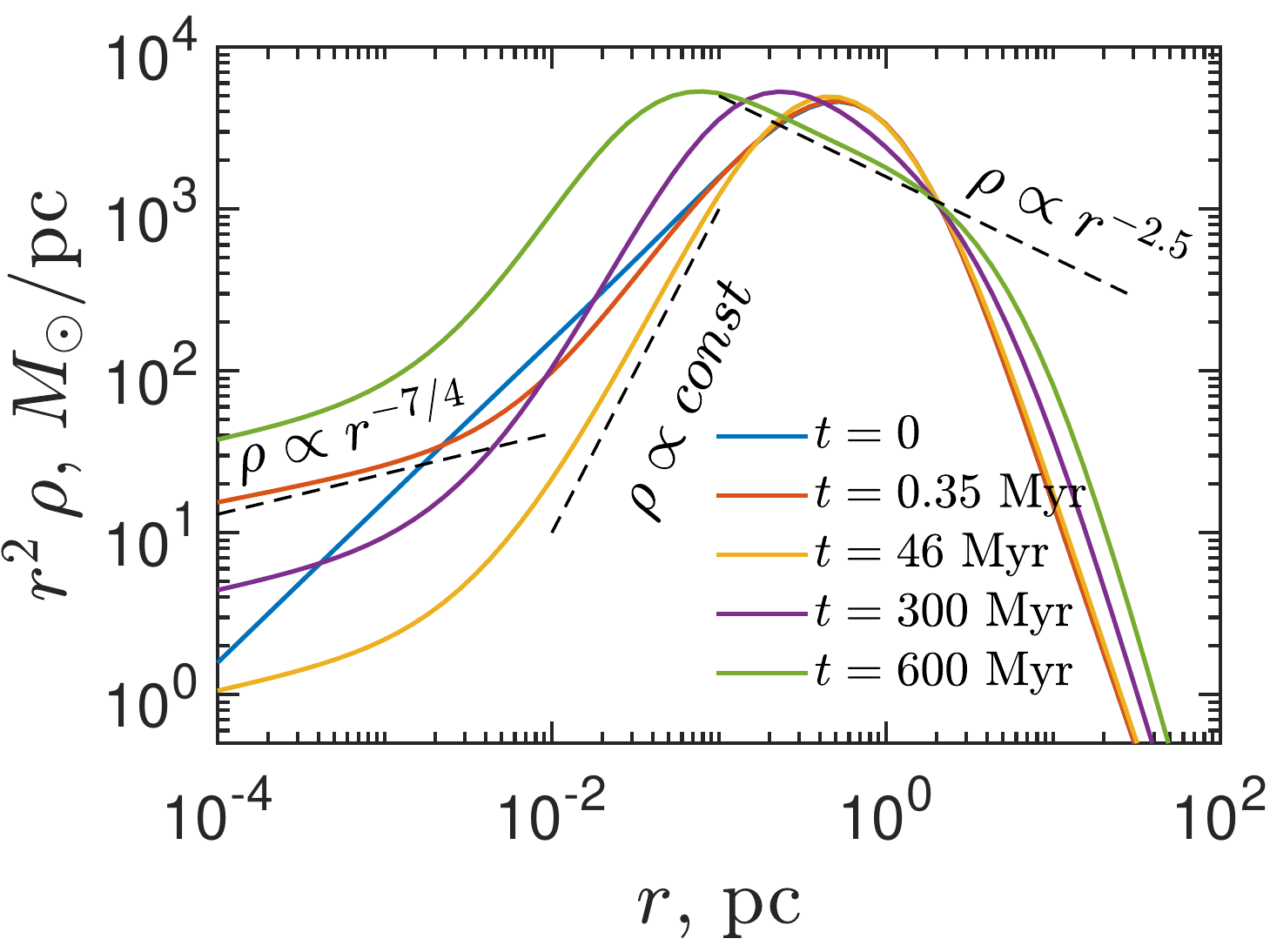}
    \caption{}
    \label{signle_pre_cc}
    \end{subfigure}
    \begin{subfigure}[t]{0.48\linewidth}
    \includegraphics[width=\linewidth]{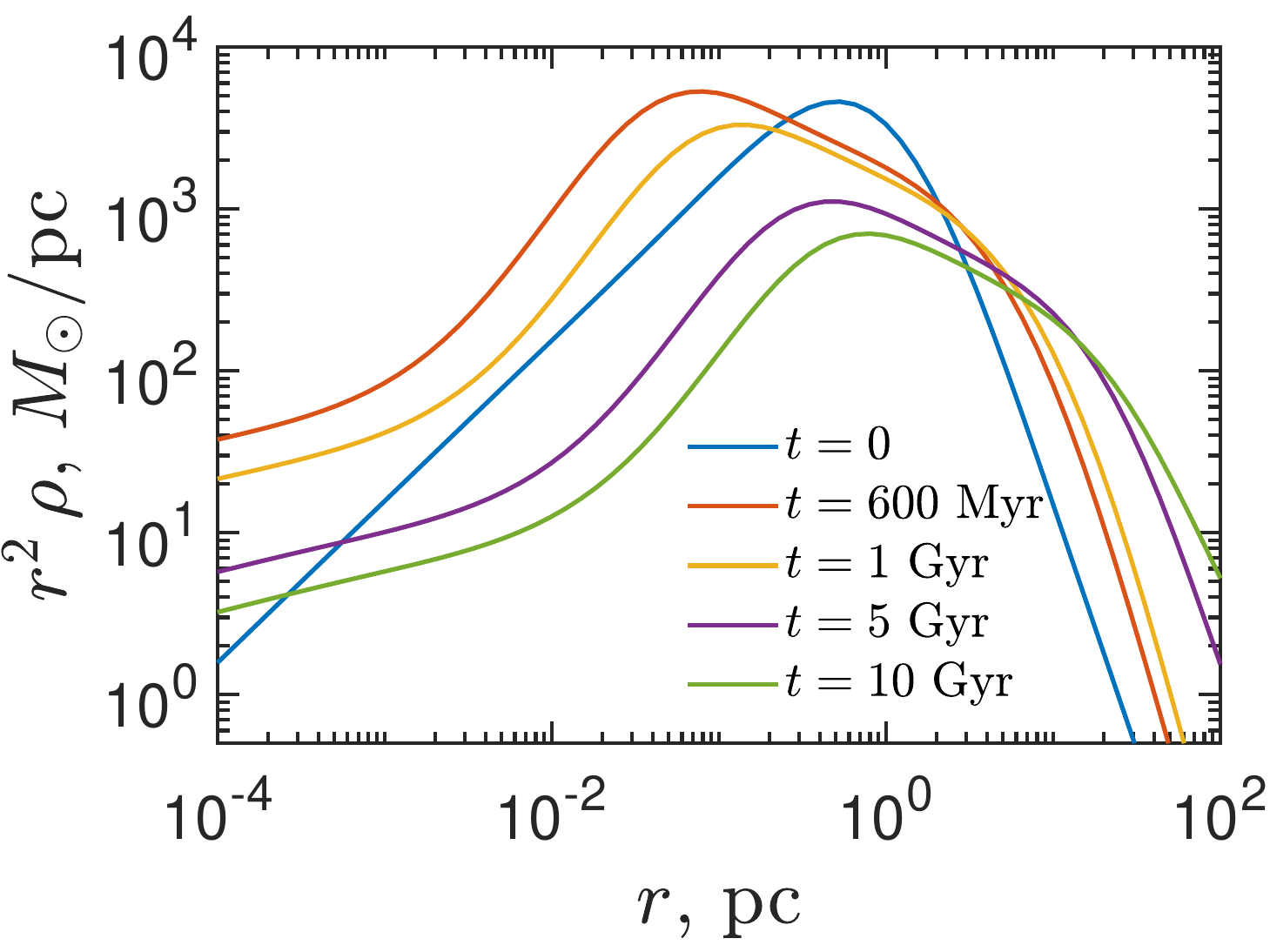}
    \caption{}
    \label{signle_post_cc_r}
    \end{subfigure}
    \caption{The evolution of the PBHs mass distribution in the cluster before (Fig.~\ref{signle_pre_cc}) and after (Fig.~\ref{signle_post_cc_r}) the self-similar expansion phase is shown.}
    \label{single_md}
\end{figure}

\begin{figure}
    \centering
    \includegraphics[width=0.5\linewidth]{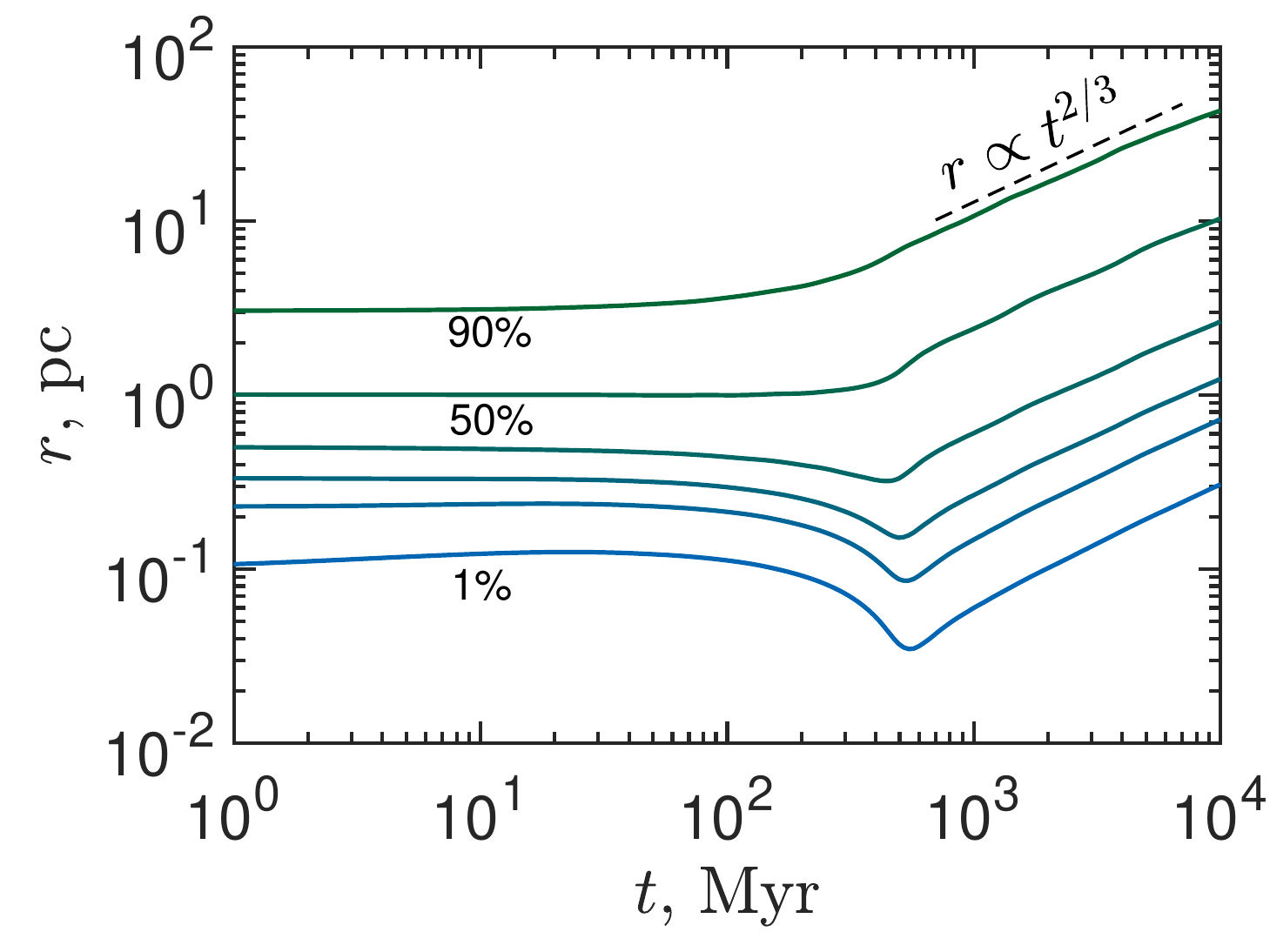}
    \caption{The evolution of mass shells is shown (the change with time of a radius which comprises a fixed fraction of the cluster mass). The curves from bottom to top correspond to 1, 5, 10, 20, 50 and 90  percent of the cluster mass, respectively.}
    \label{single_radii}
\end{figure}

The evolution of the PBHs density profile is presented in Fig.~\ref{single_md}. It should be noted, that the figure shows differential mass distribution $ \D{M(r)}/\D{r} \propto r^2 \rho(r)$. One can see, that the slope of the density profile $\rho \propto r^{-7/4}$ appears in the central region of the cluster after $\sim 0.4$~Myr. This distribution is typical for gravitationally interacting bodies around a massive black hole and is known as the cusp of Bahcall-Wolf. Then, the core is formed (yellow curve) outside the region of influence of the CBH. The further evolution follows the path of the classical gravothermal catastrophe: the inner part of the cluster is shrunk, and the central density increases. The green curve shows the final moment of the core collapse. After $\sim 600$~Myr, the stage of near self-similar expansion begins (see Fig.~\ref{signle_post_cc_r}). The evolution of mass-shells are presented in Fig.~\ref{single_radii}. One can see that after $\sim 600$~Myr all mass shells expand according to the law $r \propto t^{2/3}$ which indicates the self-similar expansion character.

\begin{figure}[!th]
    \centering
    \begin{subfigure}[t]{0.48\linewidth}
    \includegraphics[width=\linewidth]{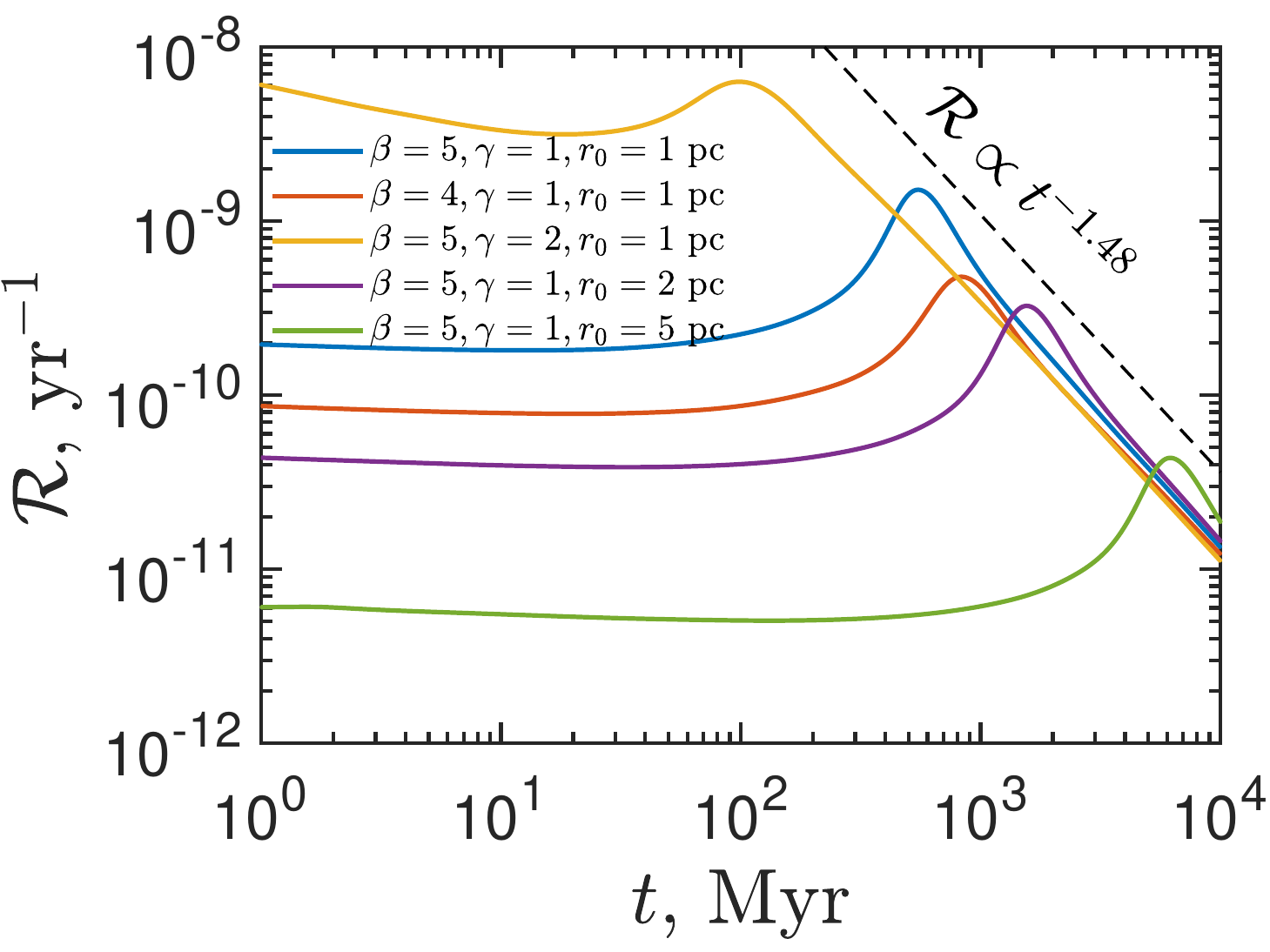}
    \caption{}
    \label{merger_rate_l}
    \end{subfigure}
    \begin{subfigure}[t]{0.48\linewidth}
    \includegraphics[width=\linewidth]{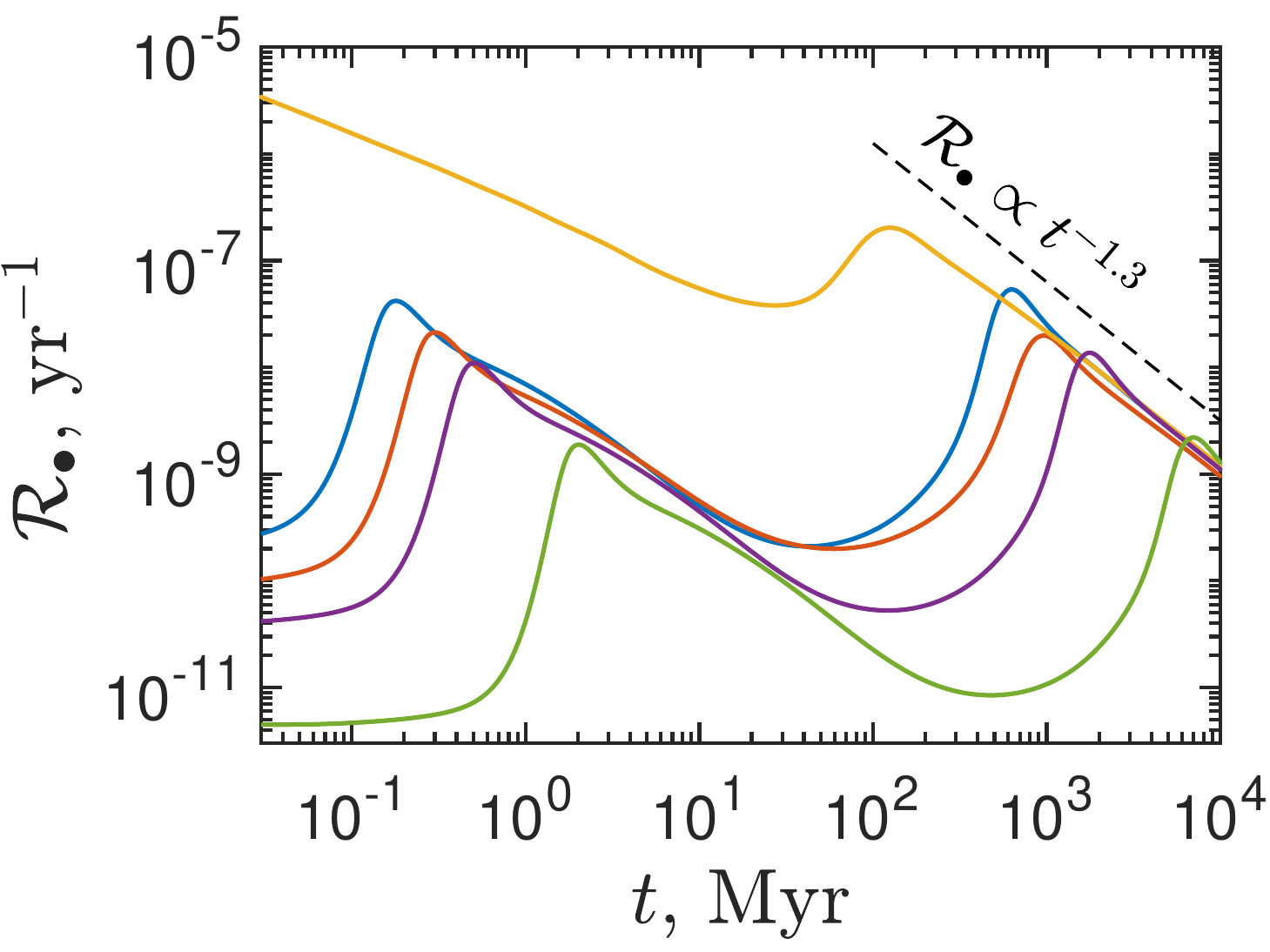}
    \caption{}
    \label{merger_rate_r}
    \end{subfigure}
    \caption{The time dependence of the merger rate of PBHs for the different parameters of the density profile is shown. Fig.~\ref{merger_rate_l}: the merger rate of PBHs between each other. Fig.~\ref{merger_rate_r}: the capture rate of PBHs by the central massive BH. The legend is the same as in the left figure.}
    \label{merger_rate}
\end{figure}

The merger rate of PBHs is presented in Fig.~\ref{merger_rate}. One can see that the modern mergers are almost independent of the initial structure of the cluster (for the considered cluster mass). It is interesting to note that after the core collapse, all curves merge into one. It can be seen that the time evolution of the merger rate has a characteristic maximum at the moment of core collapse, see Fig.~\ref{merger_rate_l}. However, for cluster with a ``steeper'' density profile at the center, this maximum is not so pronounced (the case $\gamma = 2$). In addition, it should be noted that the capture rate of CBH has two peaks, see Fig.~\ref{merger_rate_r}. The second peak is similar to the peak shown in Fig.~\ref{merger_rate_l}, while the first one is caused by formation of the Bahcall-Wolf cusp.

The asymptotic behavior of the merger rate is different for various masses of PBHs. Namely, the merger rate of approximately equal PBHs masses behaves as $\mathcal{R} \propto t^{-1.48}$. This dependence is very easy to understand due to the fact that $\mathcal{R} \propto r_c^{-31/14}$, see Eq.~\eqref{mr_2}, while all the mass shells evolve as $r \propto t^{2/3}$. On the other hand, the merger rate of black holes of significantly different masses (CBH with less massive PBHs in the cluster) evolves as $\mathcal{R}_{\bullet} \propto t^{-1.3}$. The dependence does not behave like $\propto t^{-1.68}$ (see Eq.~\eqref{cbh_mr_1}) because the mass of CBH $M_{\bullet}$ is also time dependent.

Thus, the study of the PBHs merger rate on cosmological time scales can be used to test the hypothesis of PBHs cluster existence. If future generations of gravitational waves experiments will detect the decreasing merger rate with
decreasing redshift according to the obtained dependencies, it will be an indirect evidence of the existence of PBHs clusters. Moreover, if the dependence $\mathcal{R} \propto t^{-1.48}$ is observed, it will indicate that PBHs clusters do not have a massive central BH. In this case, the expansion is associated with formation of binary systems, similar to globular star clusters. 
Note, that the clusters can have different parameters and, respectively, positions of maxima of merger rate at the redshift axes. Integration over all the clusters should change the expected merger rate dependence from $z$. 
However, the asymptotic behavior at large times will follow the obtained dependence.

\subsection{PBHs clusters with wide mass spectra}

Cluster may contain PBHs distributed by masses in a wide range as it is predicted, e.g., in the model of the domain walls collapse \cite{2019EPJC...79..246B}.
This fact leads to a shift in time scales due to mass segregation and, as a consequence, to acceleration of the core collapse. In this section, we present calculation of PBHs cluster evolution with the mass spectrum $\D{N}/\D{m} \propto m^{-2}$ \cite{2019EPJC...79..246B}
and the mass range from $10^{-2} \, M_{\odot}$ to $10 \, M_{\odot}$ with the CBH mass $100 \, M_{\odot}$ and the total cluster mass $10^5 \, M_{\odot}$. The parameters of the density profile \eqref{PBH_profile} are $\gamma = 1$, $\beta = 5$, $r_0 = 1$~pc.

\begin{figure}[!th]
    \centering
    \begin{subfigure}[t]{0.48\linewidth}
    \includegraphics[width=\linewidth]{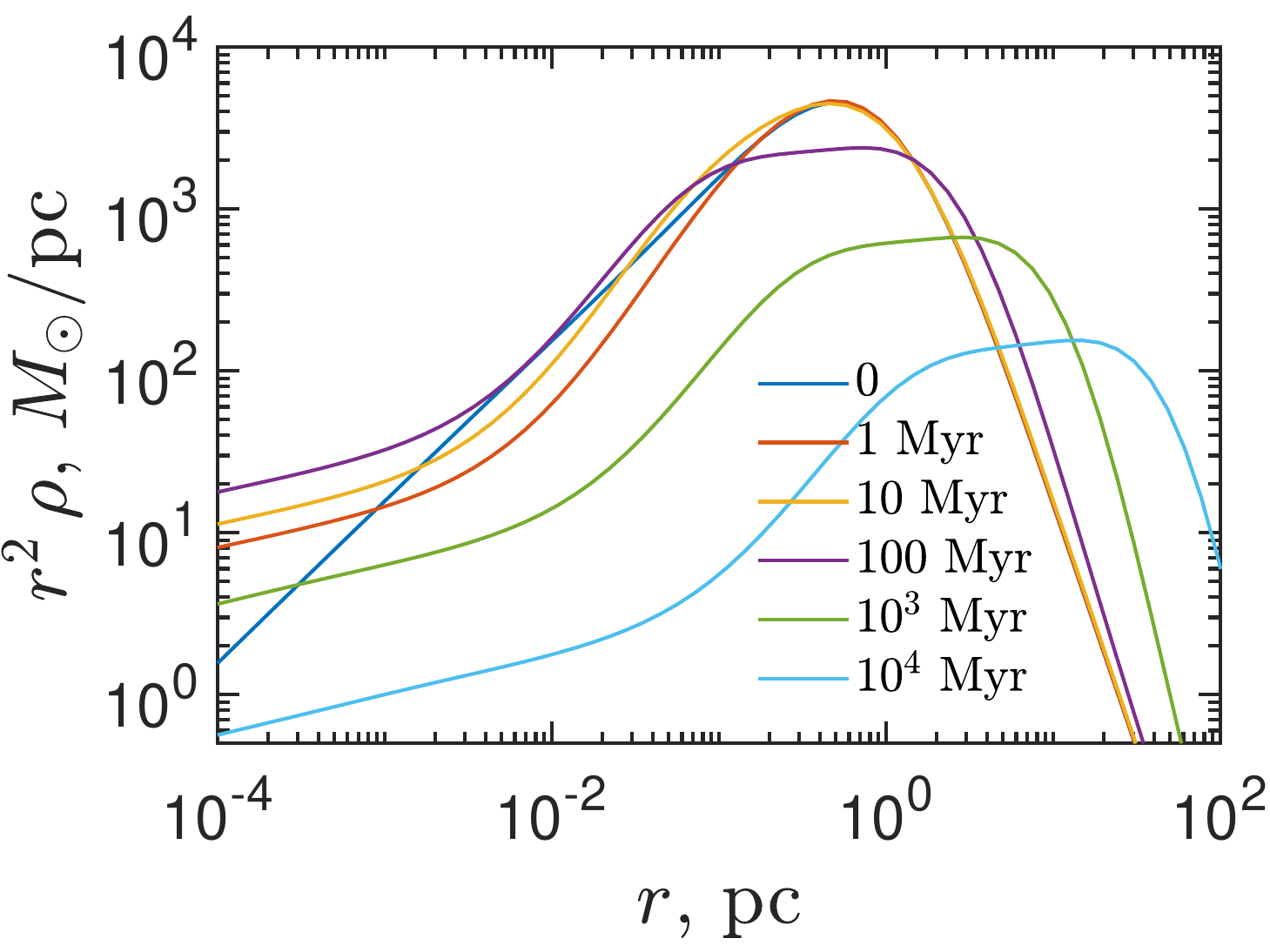}
    \caption{}
    \label{mass_distr_spectr_l}
    \end{subfigure}
    \begin{subfigure}[t]{0.48\linewidth}
    \includegraphics[width=\linewidth]{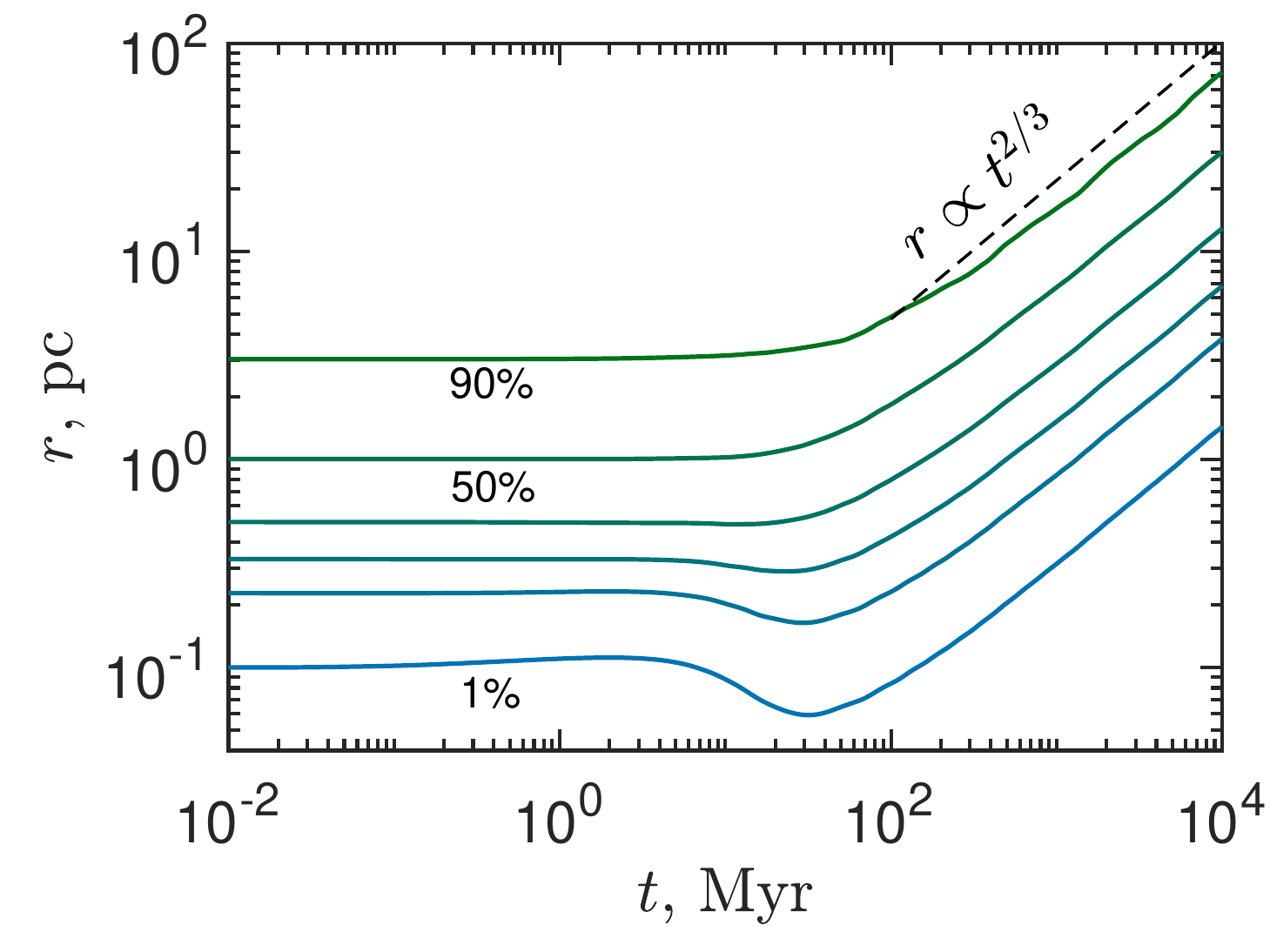}
    \caption{}
    \label{mass_distr_spectr_r}
    \end{subfigure}
    \caption{The evolution of the mass distribution is shown. Fig.~\ref{mass_distr_spectr_l}: The density profile at different times. Fig.~\ref{mass_distr_spectr_r}: The evolution of the mass shells containing 1, 5, 10, 20, 50, 90  percent of the cluster mass, respectively.}
    \label{mass_distr_spectr}
\end{figure}

\begin{figure}[!th]
    \centering
    \begin{subfigure}[t]{0.48\linewidth}
    \includegraphics[width=\linewidth]{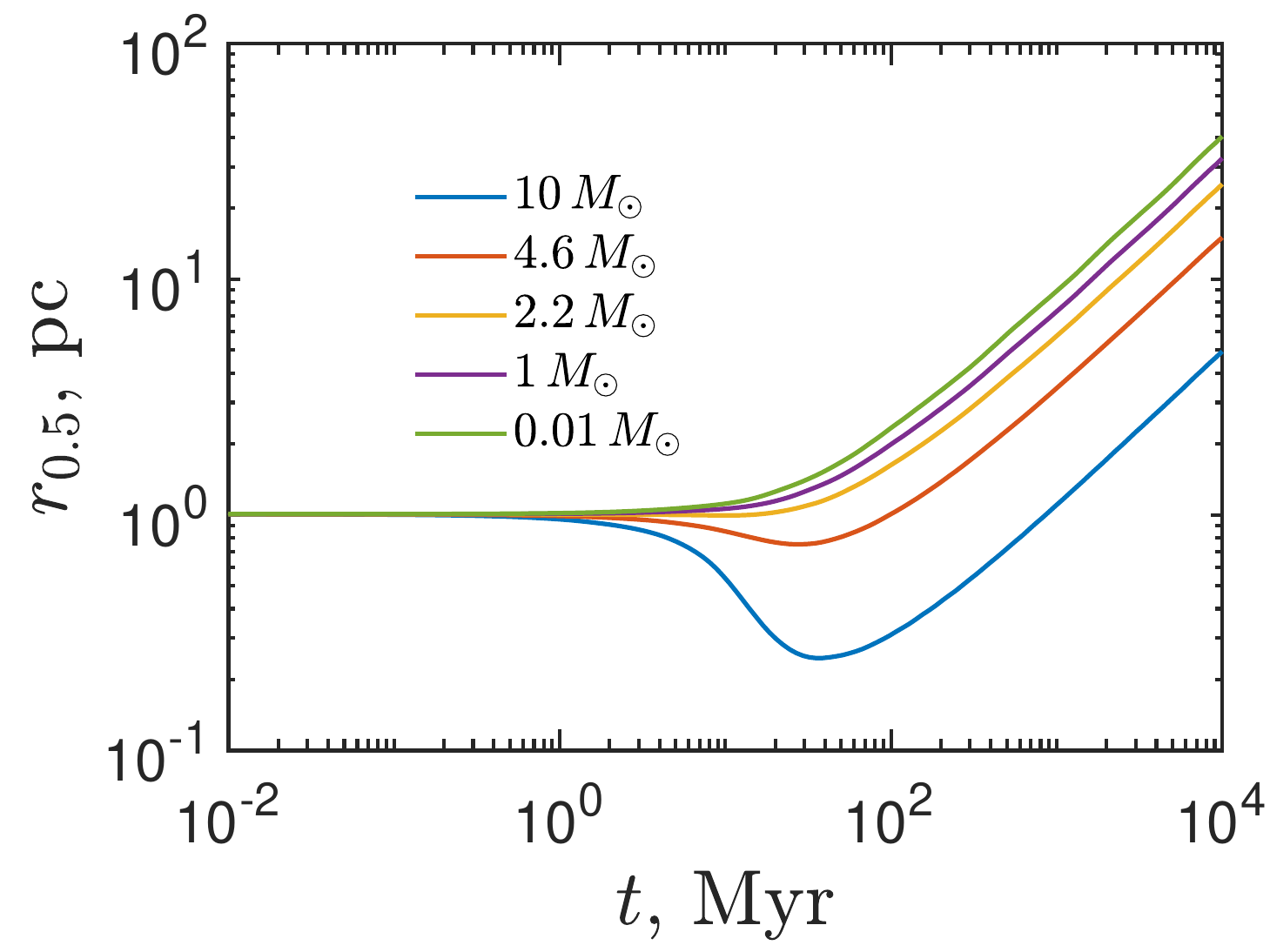}
    \caption{}
    \label{radii_diff_mass_l}
    \end{subfigure}
    \begin{subfigure}[t]{0.48\linewidth}
    \includegraphics[width=\linewidth]{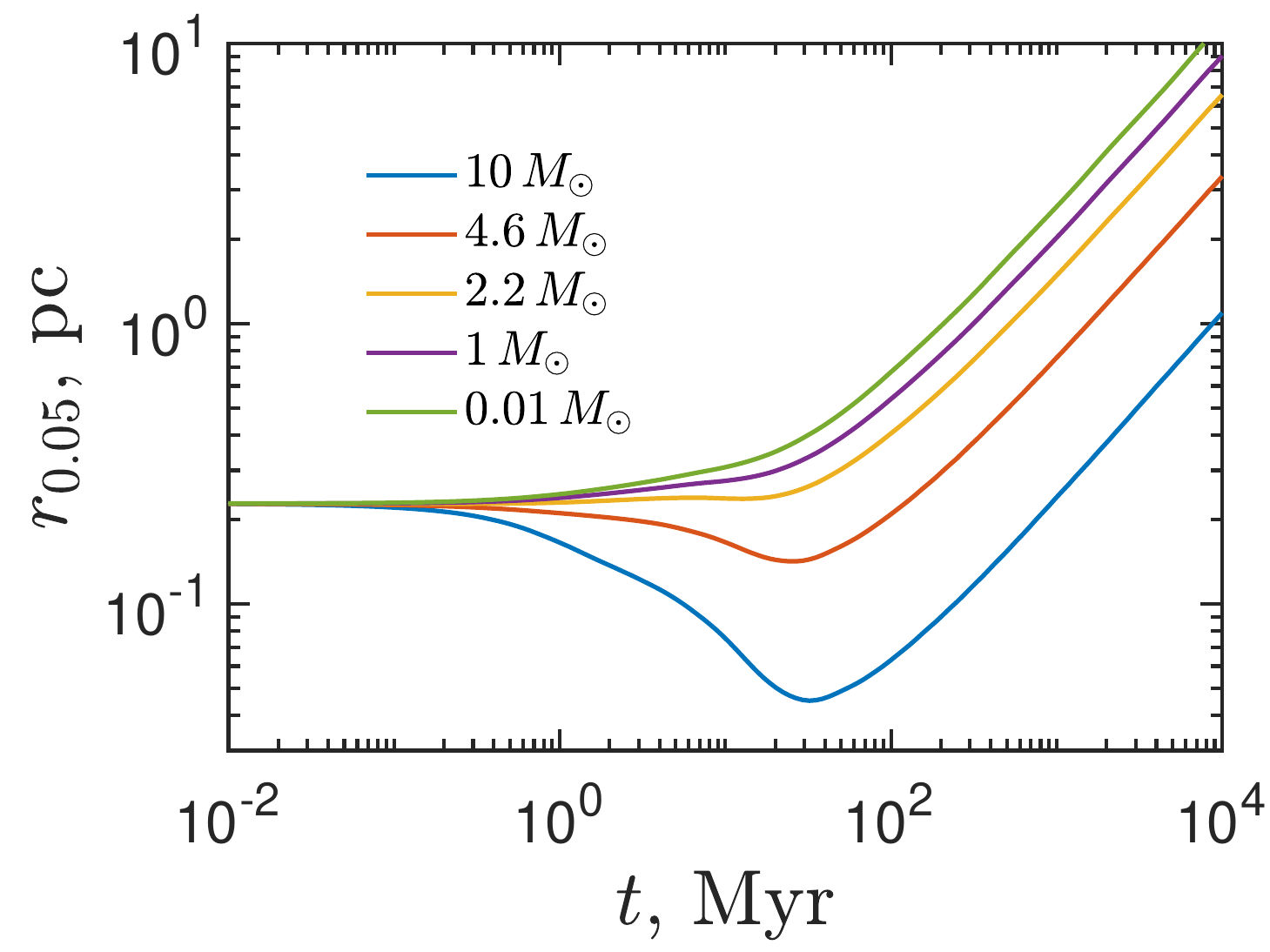}
    \caption{}
    \label{radii_diff_mass_r}
    \end{subfigure}
    \caption{The evolution of mass shells containing 50 (Fig.~\ref{radii_diff_mass_l}) and 5 (Fig.~\ref{radii_diff_mass_r}) percents of the mass for different types of PBHs is shown.}
    \label{radii_diff_mass}
\end{figure}

The evolution of the mass distribution of PBHs is shown in Fig.~\ref{mass_distr_spectr}. One can see, that the dynamics are similar to a cluster with initially monochromatic PBH mass distribution, but the density profile is near to $\rho \propto r^{-2}$. The presence of the mass spectrum of the PBHs leads to the mass segregation as result of which more massive BHs settle to the center cluster, see Fig.~\ref{radii_diff_mass}. In addition, the wide range of PBHs masses shifts relaxation timescales. The cluster enters the expansion phase after $\sim 30$~Myr while for the cluster with monochromatic PBH mass distribution this phase occurs after $\sim 600$~Myr. This fact
is caused by the formation of a subcluster from heavy PBHs (which sink in the central part of the cluster due to dynamical friction) in the central region of the ``host cluster'' which evolves faster. The expansion of the cluster is also self-similar with the law $r \propto t^{2/3}$.

\begin{figure}[!th]
    \centering
    \begin{subfigure}[t]{0.48\linewidth}
    \includegraphics[width=\linewidth]{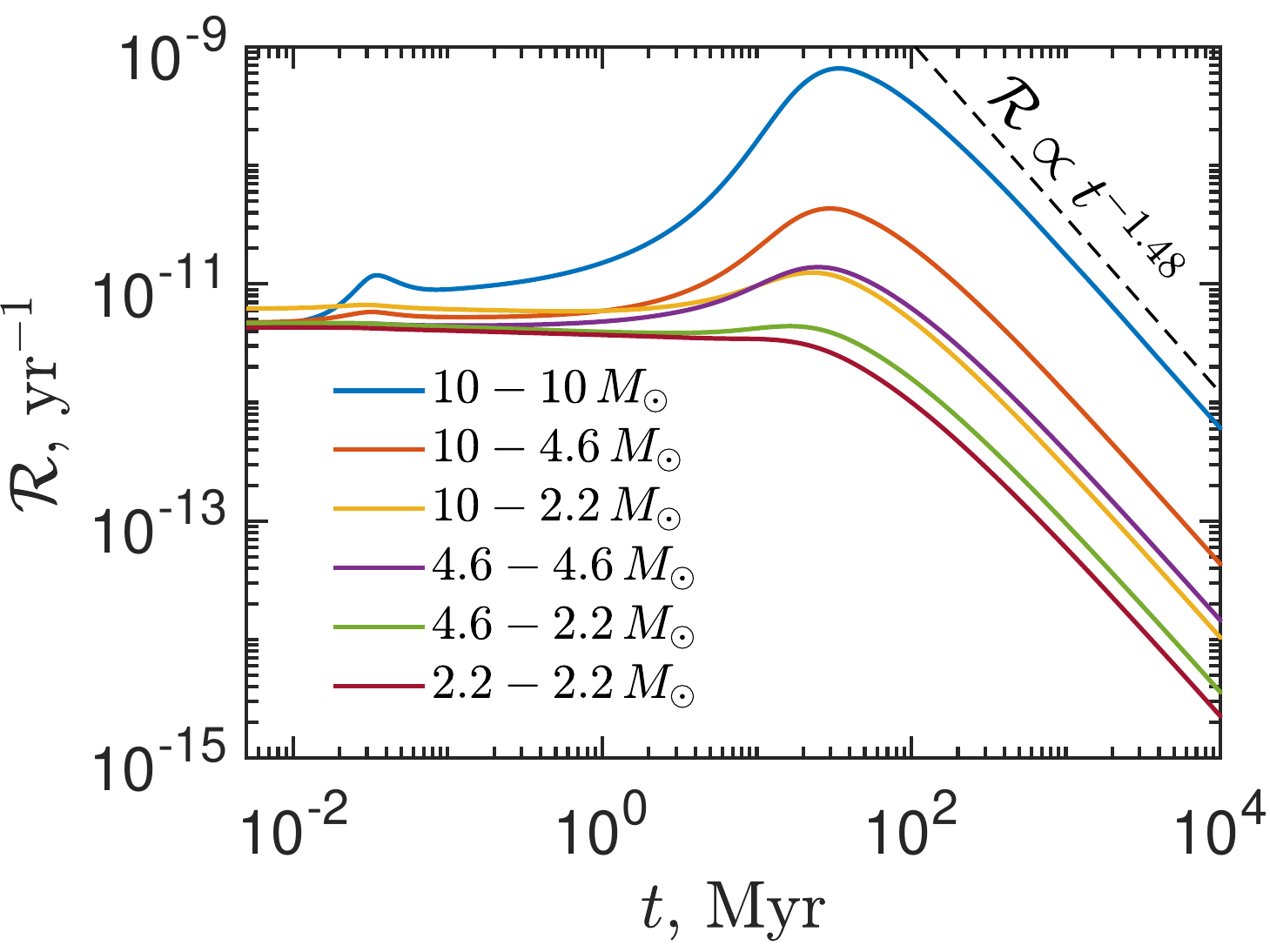}
    \end{subfigure}
    \begin{subfigure}[t]{0.48\linewidth}
    \includegraphics[width=\linewidth]{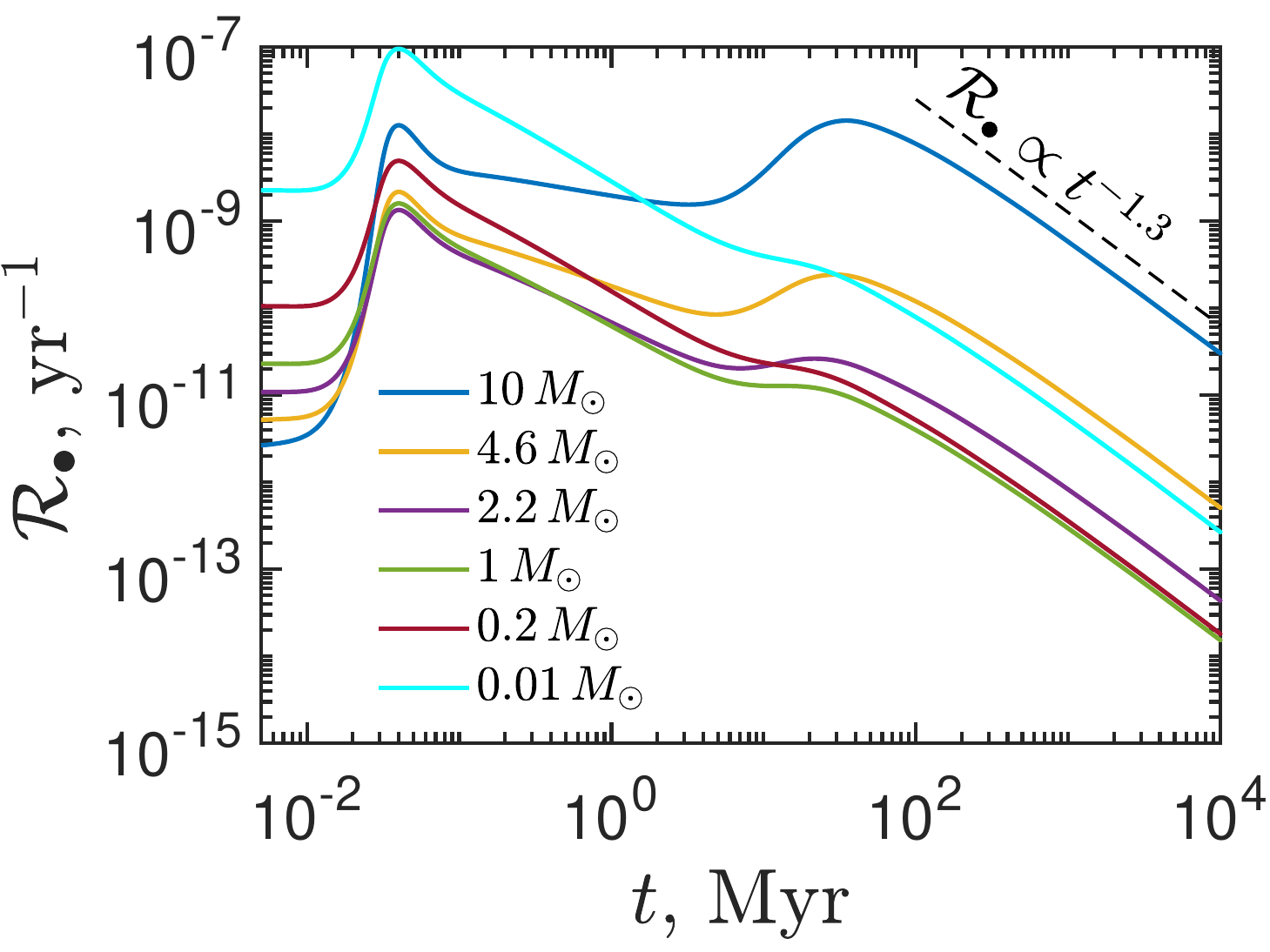}
    \end{subfigure}
    \caption{The merger rate time evolution for different type of PBHs mass. {\it{Left}}: The PBHs merger rate with each other. {\it{Right}}: The PBHs merger rate with the central massive BH.}
    \label{MergerRateSpectr}
\end{figure}

The merger rate of the different types of PBHs masses is shown in Fig.~\ref{MergerRateSpectr}.  The capture of PBHs by the central BH gives the main contribution to GWs from the cluster. However, such events are difficult to observe using modern gravitational wave detectors.

In this work, we do not take into account the external conditions for cluster. At the stage of the structure formation, clusters become a part of the DM halo. This leads to the fact that the cluster is influenced by tidal forces as a result of which it cannot expand indefinitely.
The outer layers will be captured by the gravitational potential of the host halo.
Therefore, the cluster may be partially stripped. However, as noted earlier, the obtained results are valid in order of magnitude for the central part of a cluster. Hence, even if cluster has lost most of its mass, the obtained merger rate remains correct.

One can use the obtained merger rate in order to restrict the abundance of PBHs clusters. The modern merger rate of BHs evaluated from the LIGO/Virgo data is $\mathcal{R} \sim 10$~Gpc$^{-3}$~y$^{-1}$ \cite{LIGOScientific:2020kqk} while, for instance, the merger rate for $10 \, M_{\odot}$ PBHs per the cluster is $\mathcal{R}_\text{cl} \sim 10^{-12}$~y$^{-1}$.{ From the condition $\rho_{\text{crit}} \Omega_{\text{cl}} \mathcal{R}_{\text{cl}} / M_\text{cl} \lesssim \mathcal{R}$}, one can obtain the density fraction of the PBHs cluster with the mass $10^5 \, M_{\odot}$ composed of PBHs with masses from $10^{-2}M_{\odot}$ to $10M_{\odot}$ distributed as $\mathrm{d}N/\mathrm{d}m\propto m^{-2}$ with CBH mass $100M_{\odot}$ as $\Omega_\text{cl} \lesssim 0.01$. The influence of binaries formed as a result of three-body interactions or primordial binaries formed before the formation of a cluster can have their contribution to the limit. Note that the limit on $\Omega_\text{cl}$ from GW observation can be escaped by the choice of other parameters.

\section*{Conclusion}

In this paper, the dynamics of the primordial black holes cluster was considered. We estimated the time dependence of the merger rate of PBHs in the cluster with the different parameters and showed that the cluster evolution significantly affects the merger rate. In addition, it was shown that the merger rate changes with time as $\mathcal{R} \propto t^{-1.48}$. Hence, the observation of GWs at high redshifts may shed light on the spatial distribution of PBHs in the Universe. The merger rate was considered for the PBHs cluster with both the narrow (monochromatic) and the wide mass distribution of BHs. In the first case, the merger rate has a peak at some redshift due to the core collapse. In the second case, the presence of mass distribution of PBHs inside a cluster leads to the settling of heavy BHs in the cluster center, while the lighter components tend to form outer layers of a cluster. Nevertheless, the main features of the evolution are the same as for the cluster model with the monochromatic mass distribution. We also showed that gravitational waves data constrain clusters with the total mass $10^5 \, M_{\odot}$ composed of PBHs with masses from $10^{-2}M_{\odot}$ to $10M_{\odot}$, distributed as $\mathrm{d}N/\mathrm{d}m\propto m^{-2}$ with CBH mass $100M_{\odot}$ as $\Omega_\text{cl} \lesssim 0.01$.  The last limit can be escaped by choosing other PBH cluster mass parameters. The search for similar effects as considered here or effect of gravitational microlensing on the cluster can allow testing the cluster model.

\section*{Funding}
The work of V.D.S. was supported by a grant of Russian Science Foundation No 18-12-00213-P https://rscf.ru/project/18-12-00213/ and performed in Southern Federal University (SFEDU). The work of A.A.K. was supported by the Ministry of Science and Higher Education of the Russian Federation, Project ``Fundamental properties of elementary particles and cosmology'' №~0723-2020-0041.

\sloppy
\printbibliography

\end{document}